\documentclass[a4paper,fleqn,usenatbib,referee,epsfig,color]{mnras}

\usepackage[T1]{fontenc}
\usepackage{ae,aecompl}

\usepackage{graphicx}	
\usepackage{amsmath}	
\usepackage{amssymb}	

\newif\ifAMStwofonts
\newcommand{\be}{\begin{equation}}
\newcommand{\ee}{\end{equation}}
\newcommand{\bea}{\begin{eqnarray}}
\newcommand{\eea}{\end{eqnarray}}

\newcommand{\mt}{\bf}

\title[Steady Galactic Dynamo and Halo Magnetic Fields]{\bf Steady Galactic Dynamos and Observational Consequences I: Halo Magnetic Fields }
\author[R.N. Henriksen ]
{R. N. Henriksen$^1$\thanks{henriksn@astro.queensu.ca} \\ 
$^1$Dept. of Physics, Engineering Physics \& Astronomy, Queen's University, Kingston, Ontario, K7L 3N6, Canada\\}
\date{Accepted XXX. Received YYY; in original form ZZZ}
\pubyear{2016}
\begin{document}
\label{firstpage}
\pagerange{\pageref{firstpage}--\pageref{lastpage}}
\maketitle

\begin{abstract}
We study the global  consequences in the halos of spiral galaxies of the steady, axially symmetric, mean field dynamo. We use the classical theory but add the possibility of using the velocity field  components as  parameters in addition to the helicity and diffusivity.  The analysis is based on the simplest version of the theory and uses scale-invariant solutions. The velocity field (subject to restrictions) is a scale invariant field in a `pattern' frame, in place of a full dynamical theory.  The `pattern frame' of reference may either be the systemic frame or  some rigidly rotating spiral pattern frame.  One type of solution for the magnetic field yields off-axis, spirally wound,  magnetic field lines. These predict sign changes in the Faraday screen rotation measure  in every quadrant  of the halo of an edge-on galaxy. Such rotation measure oscillations have  been observed in the CHANG-ES survey.
\end{abstract}

\begin{keywords}
Galaxies, Magnetic fields, Dynamos
\end{keywords}
\section{Introduction}
Dynamo theory as an explanation  for the magnetic fields in galaxies, including the milky way has a long history both observationally and theoretically. Recent observational reviews  and discussion of the relevant theory may be found in  \cite{Beck2015} and \cite{Kr2015}. The textbook \cite{KF2015} discusses the theory  and observation with abundant references.  \cite{Black2015} discusses modern theoretical developments . 

During the past few years the  members of the  CHANGES survey \cite{WI2015} have systematically observed the halos of edge-on galaxies using the scaled configurations of the JVLA (Jansky Very Large Array). These observations have extended our knowledge of scale heights, cosmic rays and magnetic fields in the halos of spiral galaxies.  The discovery of lagging halos (e.g. \citet{R2000},\citet{H2007}) has also led to the suggestion that the halo magnetic field is coupled to the intergalactic medium \cite{HI2016}. That model is in fact a type of $\alpha-\omega$ dynamo with diffusivity, but the $\alpha^2$ dynamo was omitted in that paper. The fields of the present paper can fill that gap.  

Over the same period increased sophistication in decoding observed rotation measures (RM) has  become common, based on rotation measure synthesis \cite{BdeB2005}, \cite{Heald2009},\cite{DamSegov2016}.  This has led to the study of rather complicated magnetic geometry in order to  connect the disc and halo magnetic fields of a spiral galaxy (see e.g. \citet{FT2014} and references, for empirical fits ). 

In this paper and the following we undertake a semi-analytic study of the magnetic fields in the discs and halos of spiral galaxies. We use the classic mean field dynamo theory based on the `$\alpha-\omega$' and the ` $\alpha^2$' terms. This is of course far from innovative, but by assuming the steady state, and scale invariance (and in this paper axial symmetry), we are able to simplify the calculations that yield the mean magnetic field. This leads to definite  global qualitative and quantitative predictions that may be compared to  observational data, such as the well-established `X' polarization pattern \cite{Kr2009}.

By `halo' we mean the halo as defined by cosmic ray particles and magnetic fields. Our approximate solutions  require us to be within several kilo parsecs of the central galactic plane, although the height increases with galactic radius. Thus at a radius of $10$ kilo parsecs, our approximations should be comfortably qualitatively correct in the region $1-3$ kilo parsecs. We do find some analytic  `toy' models that are valid at greater heights. These serve mainly to justify our approximations, including the critical rotation measure sign change in each quadrant. **This effect is related to the `mixed parity' solutions of earlier work (see discussion and references).**

 Even with the steady state assumption in axial symmetry, the full dynamo problem coupled to realistic dynamics remains a formidable theoretical problem.
 In this paper we simplify the problem further by seeking scale invariant solutions. We use a particular method that systemizes the procedure (e.g. \citet{CH1991},\cite{Hen2015}), but this is not necessary. In addition we allow the scale invariance to determine the spatial variation of the velocity field. We do not solve independently for the equations of motion so that the velocity amplitudes remain simple parameters. This allows a conveniently brief survey of the effects of various flows on the basic $\alpha^2$ and $\alpha-\omega$ turbulent dynamos and their observational consequences.

 The assumption of scale invariance is made essentially for simplicity. However it is well known  (e.g. \citet{Hen2015}) that scale invariance arises asymptotically in many physical systems, once away from initial conditions and boundaries.  This is likely to apply to the disc of a spiral galaxy beyond the bulge. Moreover the implied power law radial dependences are quite natural and flexible, if various scale invariant `classes' and different scale heights parallel and perpendicular to the galactic plane are considered.  Different `classes' are physically motivated by global conservation laws.  For example when the class $a=2$ is considered, there is a global integral of Dimension equal to that of specific angular momentum (angular momentum in a galaxy of fixed mass). Should we choose the class $a=3$, then there is a global constant of Dimension equal to that of magnetic flux. This does not automatically  exclude local sources of mean magnetic field.

 In the case of the pure $\alpha^2$ dynamo when ${\bf v}={\bf 0}$ in the pattern frame, our results are  strictly independent of the dynamics. This is also the case should ${\bf v}\parallel {\bf B}$, where both ${\bf v}$ and ${\bf B}$ are mean fields. Such a limit  allows direct contact with the halo lag model of \cite{HI2016}, wherein the only $\alpha-\omega$ effect is produced by the  differential boundary conditions between the disc and infinity. 
 The form of the helicity and diffusivity  coefficients produced by the sub-scale turbulence is also normally fixed by the assumption of scale invariance, although we suggest a generalization for subsequent investigation. We are  forced to make a `near disc' approximation when dealing with the dependence on height above the plane in our general treatment. Fortunately this is a region of observational interest.  Moreover the key result is confirmed by an analytic that holds at all heights .
 
 This paper examines axially symmetric solutions (i.e. $m=0$ in a modal analysis), but a companion paper \cite{Hen2017} uses the same methods on higher order modes based on logarithmic spirals. The total halo magnetic field must in general be a combination of the axially symmetric mode with one or more spiral modes, which introduce their own (probably weaker) quadrantal sign changes.

\section{ Formulation of Scale Invariant Equations}

We represent the magnetic field in terms of a vector potential. We ignore any conservative electric field which would require charge separation in the plasma. Then the traditional (for a steady state, we need not  worry about quenching)  dynamo equation  for the vector potential reduces to 
\be
\alpha_d\nabla\wedge{\bf A}-\eta\nabla\wedge\nabla\wedge{\bf A}+{\bf v}\wedge \nabla\wedge {\bf A}={\bf 0},\label{eq:dynamoI}
\ee
where a scaled magnetic field is given by 
\be
{\bf b}=\nabla\wedge {\bf A},\label{eq:B}
\ee
and $\alpha_d$, $\eta$ are the sub-scale `helicity' and diffusion coefficients respectively. 

 For Dimensional simplicity the scaled magnetic field is taken as  
 \be
 {\bf b}\equiv \frac{{\bf B}}{\sqrt{4\pi\rho}},\label{eq:b}
 \ee
 where $\rho$ is some fiducial density. This gives ${\bf b}$ the Dimension of velocity and so ${\bf A}$ has the Dimension of specific angular momentum.
 The formula for the vector potential requires neglecting an electrostatic field that would result from large scale charge separation.   It seems unlikely that a large scale electrostatic field should be important. Such an integration has the additional benefit of allowing the `helicity' $\alpha_d$ and the diffusion coefficient $\eta$ to vary with position without adding extra terms. 
 
 **
 Because the vector potential enters equation (\ref{eq:dynamoI}) only as the curl,  it is evident that these equations may also be written  directly in terms of ${\bf b}$. However the magnetic field equations may be written `ab initio' (see Appendix B) without making the vector potential substitution. The solenoidal condition then becomes a second constraint on the mean magnetic field.
 
 **

If we  suppose that  either ${\bf v} \parallel {\bf B}$ or that ${\bf v}={\bf 0}$, or  that the spatial dependence of ${\bf v}$ is to be given  by the scale invariance (to within arbitrary factors that appear as parameters), then these equations are complete for the magnetic field given suitable boundary conditions.  Nevertheless we discuss examples in which the presence of ${\bf v}$ is essential to the existence of a solution.

The scale invariance hypothesis requires (e.g. \citet{Hen2015}) the invariants \footnote{These are $Z$ , which fixes the conical symmetry, plus barred quantities; although $Z$ varies from cone to cone as do the quantities that depend on it.} ) and the variables  $\{r,z\}$ as
 \be
~~~~~~~~~~~~~~~~~~~~~~~~~~~~~~~~~~~~~~~~\delta r =e^{\delta R},~~~~~\delta z=Ze^{\delta R},\label{eq:rz}
\ee
and 
\bea
{\bf A}&=&\bar {\bf A}(Z)e^{(2\delta-\alpha)R},\nonumber\\
{\bf v}&=&\bar{\bf v}(Z)e^{(\delta-\alpha)R},\nonumber\\
\alpha_d&=&\bar\alpha_d(Z)e^{(\delta-\alpha)R},\label{eq:Scinv}\\
\eta&=& \bar\eta(Z)e^{(2\delta-\alpha)R},\nonumber\\
\Delta&\equiv& \frac{\bar\alpha_d}{\bar\eta\delta}.\nonumber
\eea

The parameter $\Delta$ is a kind of sub-scale Reynolds number. It is the only parameter that appears in the equations beyond the velocity field and the similarity `class' as introduced below.
 
The scale invariance is imposed by assuming that the barred quantities are independent of $R$.  
The quantities $\alpha$ and $\delta$ are arbitrary reciprocal temporal and spatial Units, and the similarity class (e.g. \citet{Hen2015}) is 
\be
a\equiv \frac{\alpha}{\delta}.\label{eq:a}
\ee
 In terms of the cylindrical variables $\{z,r\}$  the invariant $Z$ has the value 
\be
Z=\frac{z}{r},\label{eq:Z}
\ee
which is also the tangent of the angle between the radius to any point in the system and the equatorial plane.

Proceeding  with the scale-invariant analysis of equation (\ref{eq:dynamoI}) , we let 
\be
y\equiv \bar A'_r-(2-a)\bar A_z+Z\bar A'_z=\frac{\bar b_\phi}{\delta},\label{eq:y}
\ee
for notational convenience. The  magnetic field is given by ($\delta$ is the average reciprocal scale  if different vertical and radial scales are used)
\be
{\bf b}={\bar{\bf b}}e^{(\delta-\alpha)R},\label{eq:barb}
\ee
and the  components of $\bar{\bf b}$ in addition to $b_\phi$ are 
\bea
\frac{\bar b_r}{\delta}&=&-\partial_Z\bar A_\phi,\nonumber\\
\frac{\bar b_z}{\delta}&=& (3-a)\bar A_\phi-Z\bar A'_\phi.\label{eq:bfield}
\eea

The equations (\ref{eq:dynamoI}) are now reduced to three ordinary linear equations of the form
\bea
0&=&\hskip-0.75em-\bar A'_\phi\Delta +\bar u_\phi((3-a)\bar A_\phi-Z\bar A'_\phi)-y\bar u_z+y',\label{eq:Ar}\\
0&= &\hskip -0.75em (1+Z^2)\bar A''_\phi-\bar u_z\bar A'_\phi+\left((1-a)-\bar u_r)\right)(3-a)\bar A_\phi+y\Delta+\bar u_rZ\bar A'_\phi-(3-2a)Z\bar A'_\phi,\label{eq:Aphi}\\
0&=&\hskip -0.75em\bar u_\phi\bar A'_\phi+((3-a)\bar A_\phi-Z\bar A'_\phi)\Delta+y(\bar u_r-(2-a))+Zy',\label{eq:Az}
\eea
where we have set 
\be
{\bf\bar u}=\frac{{\bf \bar v}}{\bar \eta\delta},\label{eq:u}
\ee
and the prime denotes differentiation with respect to $Z\equiv z/r$. The only change in these equations if different scales in the radial and vertical directions are used is that $a\equiv \alpha/\delta=2\alpha/(\delta_\perp+\delta_\parallel)$.

These equations are reduced to ordinary equations in $Z$ because of the assumption of scale invariance. They are not commonly available  so we give them  explicitly here. They have the peculiarity of being over determined due to the scale invariance unless  the sub-scale `Reynolds number' $\Delta$  (essentially the dynamo number \citet{B2014}) is a function of $Z$. They then cease to be linear equations so that we leave them to other work. The over determination must be resolved by imposing conditions on the parameters such that two of the equations become identical.  It is important to realize that even when a  non-trivial solution may be found for an appropriate choice of parameters, the solution is not generally unique because of the over determination. Other restrictions on the parameters lead to different solutions as we show in our examples.

**

It is also useful to have the scale invariant equations  in terms of  the reduced magnetic field ${\bf b}$ (e.g. Appendix B).  Certainly if ${\bf A}$ is solved for from the three equations (\ref{eq:dynamoI}) and the  mean field  is calculated from the curl, then the divergence of the field will be zero. This is the procedure we follow normally in this paper. However we will use these equations to find exact solutions with and without zero divergence to demonstrate the sign change at large $Z$.  The physical significance of the non solenoidal examples is speculative (see appendix B  and the examples). Interestingly  there is one example  solution of these equations  ($\bar u_\phi\ne 0$) where the non solenoidal mean magnetic field  is due to the  $\alpha^2$ dynamo action itself. That is, when $\Delta\ne 0$, one encounters local sources of the mean magnetic field.
 
 **
 
 Explicitly these scale invariant, steady,  mean magnetic field  equations are:
\bea
0&=& \bar b_r\Delta+\bar u_\phi\bar b_z-\bar u_z\bar b_\phi+\bar b'_\phi,\nonumber\\
0&=& -(1+Z^2)\bar b'_r+\bar u_z\bar b_r-\bar u_r\bar b_z+\bar b_\phi\Delta+(1-a)\bar b_z+(2-a)Z\bar b_r,\label{eq:beqs}\\
0&=& Z\bar b'_\phi-(2-a)\bar b_\phi+\bar u_r\bar b_\phi-\bar u_\phi\bar b_r+\bar b_z\Delta.\nonumber
\eea

**
These equations for the magnetic field are  exact given our assumptions.  It is important to note that they are are {\it not} overdetermined even with $\Delta$ constant because of the absence of the solenoidal condition. They may therefore be readily studied numerically. If this condition is added on physical grounds, then as in the equations for the vector potential (which yield an exactly solenoidal magnetic field) they become over determined, unless $\Delta$ is a variable.

**

At this stage we remark that $\alpha_d$ and $\bar\eta$ may each be the same arbitrary function of $Z$, since the equations contain them only as $\Delta$. This allows for the same gradients of diffusivity and helicity on cones. They could be different functions if $\Delta$ were to be a function of $Z$. In that case the equations are well posed for $A_\phi(Z)$, $y(Z)$ and $\Delta(Z)$. This is an intriguing possibility, but we we will normally hold $\Delta$ constant in this paper  because otherwise the equations are non-linear.  The  reduced and scaled velocity components $\bar{\bf u}$ are also held constant, but  the scaled velocity may have the same $Z$ dependence as $\bar\eta$ (see e.g. equation (\ref{eq:u}). We note from the definition that the dependence on $Z$ is actually a dependence on $z/(r)$, so that all quantities are constant on cones but for the appropriate power law scaling factors (e.g. see equation \ref{eq:Scinv}). 


 The equations for the vector potential are difficult to solve exactly. We adopt  additional simplifying assumptions at this stage in an effort to avoid a numerical study in this exploratory paper. A simple exponential solution  is  possible if the terms explicitly dependent on $Z=z/r$ are neglected. This requires $Z$ to be small, so that the solution extends over cones relatively close to the galactic plane, although the height can easily reach a kilo parsec . 
 
 We do this everywhere  $Z$ appears in these equations, so we require both $Z$ to be small and the variation in $Z$ to be slow. Such a region `close' (say within $0.2$  of the radius of the disc) to the disc of an edge-on spiral galaxy is easily accessible observationally. However this approximation casts doubt on the physical reality of our sign change well above the plane. 
 
 **
 
 We shall nevertheless call attention to this behaviour; based in part on a smooth continuation to larger $Z$ from the small $Z$ region, and on confirmation (e.g. see figure \ref{fig:analytvphi}) provided by the solutions  that are valid for all $Z$ found in terms of the magnetic field equations (\ref{eq:beqs}).  Similar behaviour has also been suggested  and glimpsed in earlier  work (\citet{SS1990}, \cite{BDMSST92}, \cite{MS2008}) based on time dependent numerical evolution.  It is referred to as `mixed' or `intermediate' parity in those and related studies.  
 
 **
 
 Our general approach to solving for the vector potential proceeds , after neglecting the terms in $Z$ ,by solving for $y$ from equation (\ref{eq:Az}), and substituting into  equations (\ref{eq:Aphi}) and (\ref{eq:Ar}) to obtain two equations for $\bar A_\phi$.  An alternative approach involves solving for $y$ from equation (\ref{eq:Aphi}) and substituting into the other two equations. This yields a less general set of parameters for a solution, and we do not find it to be physically applicable to the observations and so omit it in the interests of brevity. In either case the two equations for $\bar A_\phi$ must be reconciled by requiring the coefficients of $\bar A_\phi$ and its derivatives  in the two equations to be identical. This imposes conditions on $\Delta$ and on the velocity components in order that a non-trivial solution exist, much as does setting the determinant of the coefficients  of a set of homogeneous linear equations to zero.

 Proceeding  in this way  the general case is reduced to one equation for $\bar A_\phi$ namely  
\be
\bar A''_\phi-(\bar u_z+\frac{\bar u_\phi}{\tilde u_r}\Delta)\bar A'_\phi-(3-a)(\tilde u_r+\frac{\bar u_z}{\bar u_\phi}\Delta)\bar A_\phi=0,\label{eq:Aeq}
\ee
where 
\be
\tilde u_r\equiv \bar u_r-(2-a).\label{eq:tildeu},
\ee
This allows simple analytic solutions.

The conditions required to obtain a non-trivial solution by equating the coefficients  of the first and second members of the scaled vector potential equations are  ($\bar u_\phi\ne 0$)
\bea
\tilde u_r^2+(3-a)\tilde u_r+\bar u_\phi^2&=&0,\label{eq:C1}\\
\Delta^2-\frac{\tilde u_r\bar u_z}{\bar u_\phi}\Delta+\tilde u_r&=&0.\label{eq:C2}
\eea
 The solution for the scaled magnetic field  is completed by the equations (\ref{eq:bfield}) and 
\be
y=-\frac{\bar u_\phi\bar A'_\phi+(3-a)\bar A_\phi\Delta}{\tilde u_r}\equiv \frac{\bar b_\phi}{\delta}.\label{eq:bphi}
\ee

We see  from equation (\ref{eq:C1}) that $\tilde u_r<0$ (assuming $a<3$ as is normally the case in this section) for a real solution.  In fact this equation requires for real $\bar u_\phi$ that 
\be 
-(3-a)<\tilde u_r<0,\label{eq:radcon}
\ee
which is a useful condition on the radial velocity.  The analysis of various cases requires some care, and two specific examples are discussed in the appropriate section.

\section{ Examples}

\subsection{Exact solutions for the mean magnetic field}

**

We turn to  analytic solutions based on equations (\ref{eq:beqs}). The first example is an exact solution to the dynamo equations for all $Z$ and it is solenoidal when the $\alpha^2$ dynamo action is absent ($\Delta=0$).
 This first example follows from equations (\ref{eq:beqs}) when only $\bar u_\phi\ne 0$. One finds that 
\bea
\bar b_r&=&\frac{Z\bar u_\phi-\Delta}{\Delta^2+\bar u_\phi^2}\bar b'_\phi,\nonumber\\
\bar b_z&=&-\frac{\bar u_\phi+Z\Delta}{\Delta^2+\bar u_\phi^2}\bar b'_\phi,\nonumber\\
0&=& (1+Z^2)(\Delta-Z\bar u_\phi)\bar b''_\phi+(\Delta-Z\bar u_\phi)Z\bar b'_\phi+\Delta(\Delta^2+\bar u_\phi^2)\bar b_\phi.\label{eq:buphi}
\eea
The equation for $b_\phi$ is more difficult in this case so we do not study it. Our purpose is to note that now the solenoidal condition is equal to 
\be
-\frac{1}{\Delta^2+\bar u_\phi^2}(\bar u_\phi(1+Z^2)\bar b''_\phi+(\Delta+Z\bar u_\phi)\bar b'_\phi)=0.
\ee
This will only hold (given a solution of the third equation in \ref{eq:buphi}) if $\Delta =0$. That is, when there is no $\alpha^2$ dynamo action. In this example it seems that the sub-scale  magnetic field generation requires a source in the mean magnetic field. 

The solution with $\Delta=0$ is a pure $\alpha-\omega$ , solenoidal,  mean magnetic dynamo (in the pattern frame) and has the solution
\bea
\bar b_r&=&\frac{Z\bar b'_\phi}{\bar u_\phi},\nonumber\\
\bar b_\phi&=& C\sinh^{-1}(Z)+\bar b_\phi(0),\label{eq:solbuphi}\\
\bar b_z&=&-\frac{\bar b'_\phi}{\bar u_\phi},\nonumber
\eea
where $sinh^{-1}(Z)$ may be expressed as  $sinh^{-1}{Z}=\ln{|Z+sgn(Z)\sqrt{(1+Z^2)}|}$ for application on both sides of the disc.  The projected linear polarization is as usual, the angle with the plane $\lambda$ being given by  
\be
\lambda =\arctan(\frac{b_z}{b_r})=-\frac{r}{z},\label{eq:XF1}
\ee
that is 
\be
tan(\pi/2-\lambda)=Z,\label{eq:XF2}
\ee

This yields projected magnetic field lines at a fixed angle to the axis above the plane, which is rather like  the observed fields (e.g. \cite{Kr2015}). There is also in this pure $\alpha-\omega$ dynamo the sign changing  (or parity changing) effect that represents the main content of our approximate solutions below, so we study it in some detail. 

\begin{figure}
\begin{tabular}{cc} 
\rotatebox{0}{\scalebox{0.4} 
{\includegraphics{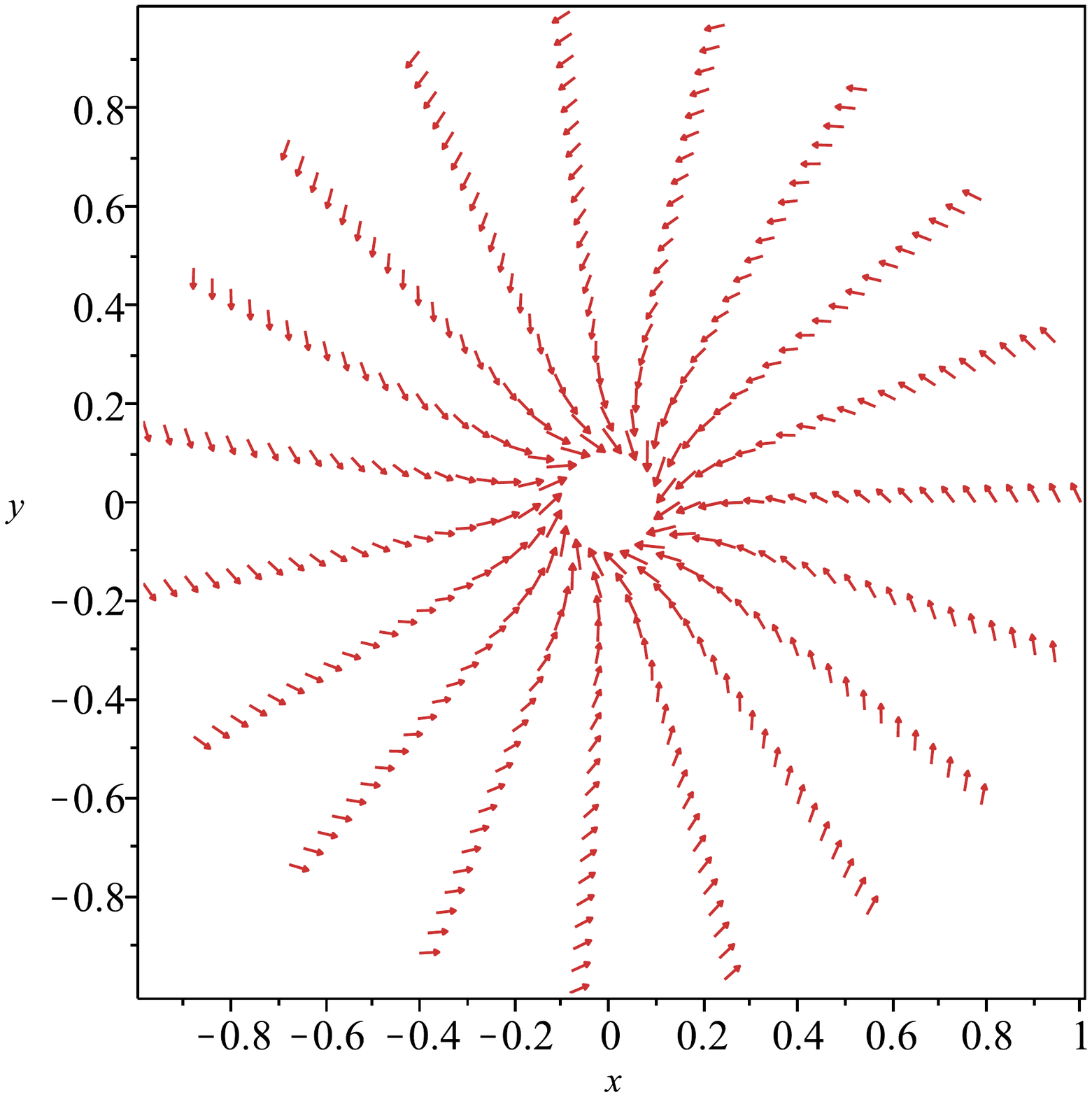}}}&
\rotatebox{0}{\scalebox{0.6} 
{\includegraphics{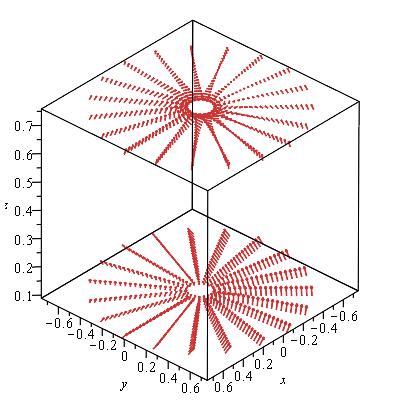}}}\\
{\rotatebox{0}{\scalebox{0.4} 
{\includegraphics{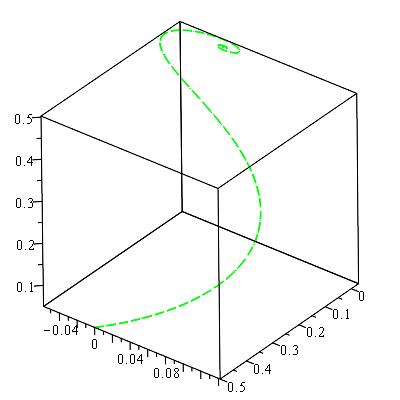}}}}&
\rotatebox{0}{\scalebox{0.4} 
{\includegraphics{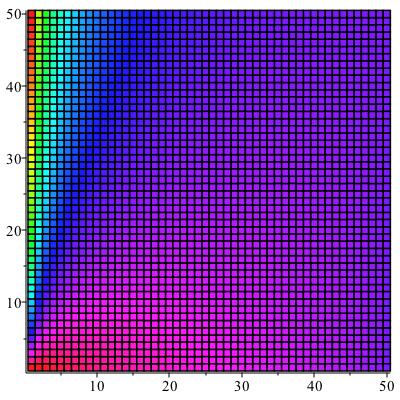}}}
\end{tabular}
\caption{At upper left we show a cut through the pure $\alpha-\omega$ exact solenoidal solution at $z=0.35$. We have set $C=-\bar b_\phi(0)=-1$. The radius runs from $0.15$ to $1.0$. At upper right there is a 3D plot of the same solution wherein each of $r$ and $z$ run from $0.15$ to $0.75$.  At lower left we show a field line from the same solution that originates in the disc at  $r=0.5$, $\phi=0$ and $z=0.05$ and continues to positive $z$. It spirals into and crosses $r=0$ near $z=0.5$ and is then reproduced on the other side of the axis, forming a closed twisted loop. At lower right we show the Faraday screen rotation measure in the first quadrant for this solution. We take $\bar u_\phi=1$ in all images except at upper right where $\bar u_\phi=2$.   }
\label{fig:analytvphi}
\end{figure}

In figure (\ref{fig:analytvphi}) we show at upper left a cut through the solution at $z=0.35$. The sign change in the azimuthal component of the field as the radius runs over the disc is evident. This is emphasized at upper right in three dimensions. We see the sign change there at the same radius but at different heights. We have also increased $\bar u_\phi$ to $2$ in this image to show the consequent  increase in the azimuthal field  relative to the radial field. 

At lower left a field line is shown spirally converging to the axis. This field line would cross the axis and be reproduced on the other side, thus forming a large (it reaches $z=0.5$) loop over the galactic centre. Because of the axial symmetry, this will appear at all azimuthal angles.  The off axis twisting of the field line is the origin of the sign changing effect that we find in our approximate solutions. 

At lower right we show the Faraday screen (see the general discussion below) rotation measure in the first quadrant. The reddish to reddish -blue colour is generally positive , while the dark blue to green to yellow or orange is generally negative. Of course the signs may be reversed as there is an arbitrary constant amplitude. It is also not a numerical fit to observational data but indicates the qualitative behaviour that may have been found  observationally \cite{CMP2016}.

The solution gives the azimuthal field to increase  logarithmically  but indefinitely with height, while the radial field becomes constant and the vertical field vanishes. As such the solution can not be continued into the inter galactic medium, but may indicate increased strength of the field in the halo. 
The general importance of this example is mainly to confirm  at large $Z$ the  existence of the sign-changing effect that we infer in our approximate solutions .

In order to demonstrate that non solenoidal solutions also follow from equations (\ref{eq:beqs}) we proceed to give an elegant example. Its physical significance is not clear (see a brief speculation in Appendix B and below), but it exhibits many of the same characteristics found in our solenoidal examples. 

We assume either ${\bf v}={\bf 0}$ or ${\bf v}\parallel {\bf b}$ and choose the scale invariant class $a=2$. 
 This choice implies a globally conserved angular momentum in a galaxy of fixed mass. The velocity and magnetic fields vary in cylindrical radius $r$ as $r^{-1}$. The helicity and diffusivity are not necessarily constant on cones so long as $\Delta$ and ${\bf u}$ are.
 Then equations (\ref{eq:beqs}) yield 
\bea
\bar b_r&=&-\frac{\bar b'_\phi}{\Delta},\nonumber\\
\bar b_z&=& -\frac{Z\bar b'_\phi}{\Delta},\label{eq:balt}\\
0&=&(1+Z^2)\bar b''_\phi+Z\bar b'_\phi+\Delta^2\bar b_\phi,\nonumber
\eea
and the last equation has the solution
\be
b_\phi=C1\sin{(\Delta\ln{(|Z+sgn(Z)\sqrt{1+Z^2}|)})}+C2\cos{(\Delta\ln{(|Z+sgn(Z)\sqrt{1+Z^2}|)})},\label{eq:bphialt}
\ee
where we have again written $sinh^{-1}{Z}=\ln{|Z+sgn(Z)\sqrt{(1+Z^2)}|}$ .
One can note that the expected projected linear polarization in this type of solution is at an angle $\lambda =\arctan {(b_z/b_r)}$ to the galactic disc. In the solution above this is simply equal to $\arctan{(Z)}$, which makes the projected polarization X-type (i.e. diverging from the galactic plane). The  magnetic field  oscillates slowly but indefinitely at large $Z$ and thus like the solenoidal solution can not be continued into the intergalactic medium.  The oscillation is however reminiscent of the temporal azimuthal field oscillations found in numerical work (e.g. \citet{MS2008}). 

This is an exact  solution to the axially symmetric $\alpha^2$ dynamo equations (possibly with parallel velocity and magnetic fields),  which also contains a possible sign changing effect, but the divergence is not zero. In our variables zero divergence requires 
\be
-Z\bar b'_r+\bar b'_z=0.\label{eq:div1}
\ee
Inserting the values for $\bar b_r$ and $\bar b_z$ from equations (\ref{eq:balt}) shows that the last expression becomes equal to $\bar b_r\equiv rb_r$.  

 We recall that we solve for a mean field, which  we speculate  (see end of Appendix B) may physically have a non-zero divergence if the mesoscale averaging volume  varies with position. We further speculate that this  may happen near a physical boundary such as the galactic disc, because the turbulent intensity can vary rapidly and anisotropically there.  This would change the volume over which a constant $\alpha_d$ may be calculated.

**

Our final example using equations (\ref{eq:beqs}) illustrates another choice of similarity class, in which the magnetic field is solenoidal. We used $a=2$ above and will use $a=1$ below. Here we  take $a=3$ , which implies  a global constant with the Dimensions of magnetic flux\footnote{$\ell^2 b$ has the Dimensions $\ell^3/t$, hence $\alpha=3\delta$ or $a=3$.}. The solenoidal condition  becomes now 
\be
-Z\bar b'_r+\bar b'_z-\bar b_r=0.\label{eq:div2}
\ee
 Using equations (\ref{eq:beqs}) we see by inspection that if we take $\bar u_r=-1$, then we have 
 \bea
 \bar b_r&=&-\frac{\bar b'_\phi}{\Delta},\nonumber\\
 \bar b_z&=&-\frac{Z\bar b'_\phi}{\Delta},\label{eq:solbur}\\
 0&=& (1+Z^2)\bar b''_\phi+2Z\bar b'_\phi+\Delta^2\bar b_\phi,\nonumber
 \eea
 and equation (\ref{eq:div2}) is satisfied. The projected linear polarization is  X-type with $\lambda=\arctan{(Z)}$ . The solution to the equation for $\bar b_\phi$ is 
 \be
 \bar b_\phi=C_1P_{arg(\Delta)}(iZ)+C_2Q_{arg(\Delta)}(iZ),\label{eq:solbphiflux}
 \ee
 where 
 \be
 arg(\Delta)=\frac{\sqrt{1-4\Delta^2}-1}{2},\label{eq:arg}
 \ee
and $P$ and $Q$ refer to the Legendre functions of the first and second kinds. We recall that the magnetic and velocity fields will have the above forms multiplied by $r^{-2}$ in this case.   

\begin{figure}{h}
\begin{tabular}{cc} 
\rotatebox{0}{\scalebox{0.6} 
{\includegraphics{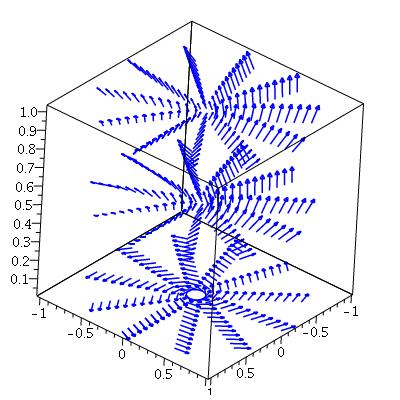}}}&
\rotatebox{0}{\scalebox{0.4} 
{\includegraphics{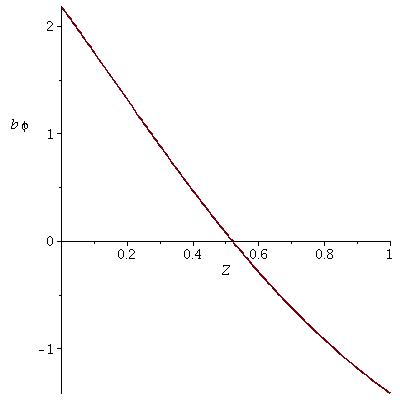}}}
\end{tabular}
\caption{ On the left we show a vector plot of the $a=3$ flux conserved solution. We have set $\Delta=0.5$, and $C=0$ and $\bar u_r=-1$.The radius of the disc is one Unit, which is also the Unit in $z$. On the right we have plotted $\bar b_\phi$ for the case $\Delta=1.0$, $C=1.0$  and $\bar u_r=-1$. There is a substantial sign reversal in $\bar b_\phi$ as a function of $Z$. This leads to a reversal in the rotation measure on the same side of the minor axis.  }
\label{fig:fluxfig}
\end{figure}

We indicate only two aspects of this solution in figure (\ref{fig:fluxfig}). On the left we show a field vector plot of one example of this solution for comparison with the similar plot in figure(\ref{fig:analytvphi}). The similarity is striking.  On the right of the figure we show a case where the sign of the azimuthal field changes dramatically with increasing $Z$ in each quadrant.  There must be off axis spirals as we indeed find below. It is one of our main observable results.

\subsection{General Resolution of Over Determinateness}

In this section we pursue the general approximation with $\Delta$ constant as summarized in equations (\ref{eq:Aeq}),(\ref{eq:bfield}),(\ref{eq:bphi}) and (\ref{eq:C2}),(\ref{eq:C1}).

We solve the linear equation (\ref{eq:Aeq}) in the general form 
\be
\bar A_\phi=C^+_\phi e^{p_+Z}+C^-_\phi e^{p_-Z},\label{eq:Asolexp}
\ee
where $p$ and $C_\phi$  are complex in general. The equation for the two values of $p$ becomes 
\be
p=\frac{1}{2}(\bar u_z+\frac{\bar u_\phi}{\tilde u_r}\Delta)\pm \frac{1}{2}\sqrt{(\bar u_z+\frac{\bar u_\phi}{\tilde u_r}\Delta)^2+4(3-a)(\tilde u_r+\frac{\bar u_z}{\bar u_\phi}\Delta)}.\label{eq:pval}
\ee
 The conditions (\ref{eq:C1}) and (\ref{eq:C2}) are usefully written as 
 \be
 \bar u_\phi=\pm\sqrt{-\tilde u_r^2-(3-a)\tilde u_r},\label{eq:C11},
 \ee
 and 
 \be
 \Delta=\frac{1}{2}\frac{\tilde u_r\bar u_z}{\bar u_\phi}\pm\frac{1}{2}\sqrt{(\frac{\tilde u_r\bar u_z}{\bar u_\phi})^2-4\tilde u_r},\label{eq:C22}
 \ee
 particularly when we remember condition (\ref{eq:radcon}).
 Whenever $p$ is complex we write the solution for $\bar A_\phi$ in the form 
 \be
 \bar A_\phi=C_\phi \exp{(Re(p)Z)}\cos{(Im(p)Z+\Phi)},\label{eq:Asolosc}
 \ee
 rather than the  single or double exponential that applies when one or both values of $p$ are real and negative ($Z>0$).

 One can  in general calculate a simple expression for the angle of the projected linear polarization namely $tan^{-1}(\bar b_z/\bar b_r)$ for these solutions. In the case of complex $p$ this becomes from equations (\ref{eq:bfield})
 \be
 tan^{-1}(\frac{\bar b_z}{\bar b_r})=\arctan{(Z-\frac{(3-a)}{ (Re(p)-Im(p)tan(Im(p)Z+\Phi))})}.\label{eq:projpol}
 \ee
It becomes $\arctan{(Z)}$ for all cases if $a=3$ as in our last example of the previous section. In general, $a$ and the various velocities would have to be fit to the observations. 


 Although the  inferred fields are more complicated than either a pure dipole or  a pure quadrupole, it is convenient to label a  magnetic field with zero vertical field at the disc ($Z=0$) as `quadrupolar', while that with a non-zero vertical field may be said to be `dipolar'.  This is not quite the same as these designations in \cite{KF2015}, where even symmetry on crossing  the disc is quadrupolar and odd symmetry is dipolar.  Our solutions  show  mainly mixed symmetry in the tangential field components on crossing the disc. 
 
 Equation (\ref{eq:bfield})  together with equation (\ref{eq:Asolosc}) shows that the quadrupolar boundary condition is applied when $\Phi=\pi/2$.  A numerical example requires the choice of $\tilde u_r$, $\bar u_z$, $\Phi$, $C_\phi$ and $a$ as parameters. Until we are concerned with observed quantities, we may set $C_\phi=\pm 1$. Moreover the similarity class $a$ is normally fixed by some physical quantity that we wish to conserve under the rescaling operation. For example if we work in the systemic reference frame where there is a constant disc  velocity, then we require the homothetic  similarity class with $a=1$.  Consequently, since $\Phi$ determines essentially the boundary condition, we are left with $\tilde u_r$ and $\bar u_z$ as principal physical parameters, with $\tilde u_r$  nevertheless  restricted by the condition (\ref{eq:radcon}).   
 
 In addition there are three `switches'  $s1$,$s2$ $s3$ , each having the values $\pm 1$ that correspond to the signs chosen  for the radicals in the expressions for $\bar u_\phi$, $\Delta$ and $p$ respectively. It transpires that these switches play an important role in allowing $\bar u_z$ to change sign across the disc while maintaining the same numerical solution on each side. We proceed in the next section with an  example of this type.
 
 \subsection{The Halo Magnetic Field}.
 
 We choose $\tilde u_r=-0.9$ and $\bar u_z=\pm 0.3$ at positive or negative $z$ respectively.  We will suppose a constant disc  velocity  in the systemic reference system so that a consistent similarity class is $a=1$ .  We can drop the bar over the velocity components and over the magnetic field components  as a result since  this implies $\alpha=\delta$.  We are proceeding with the isotropic scaling rather than the anisotropic possibility discussed in the Appendix. This choice of $\tilde u_r$ gives $u_r=0.1$. We have chosen $s1=-1$ which gives $u_\phi=-0.995$ so that the disc is rotating according to the left-hand rule (positive $z$ axis up). This is rather arbitrary relative to right/left line of sight disc velocity, since this may be reversed by inverting the solution. The number being close to unity suggests that our velocity scale is set by the disc rotational velocity. In fact the velocity scale is set through $\bar\eta\delta\equiv \eta/r$ (see equation \ref{eq:u}), which requires in this case  $\eta/r\approx 200$ km/sec.  
 
 The vertical outflow is thus $30\%$ of the value set by $\bar\eta\delta$, while the disc velocity is $99\%$ of this velocity. The switch $s2$ is  set =$-1$  for $z>0$ and $u_z>0$ but  becomes $+1$ below  the plane. The switch $s3$ also changes sign on crossing the plane. We take $C_\phi=1$ , but this is arbitrary. These guarantee that $\bar b_z$ , $\bar b_\phi$ are continuous across the plane, although $\bar b_r$ changes sign. There is no change in the handedness of the field on crossing the plane.
  When  $\Phi=\pi/2$ the vertical field at the disc is zero, which is the quadrupolar special case.
 
   There is a complementary boundary condition, found  by changing the sign of $C_\phi$  crossing the disc  and leaving the switch changes the same. Then  $b_\phi$  changes sign but $b_r$ does not . The handedness of a field line is  changed in that case.

\begin{figure}
\begin{tabular}{cc} 
\rotatebox{0}{\scalebox{0.4} 
{\includegraphics{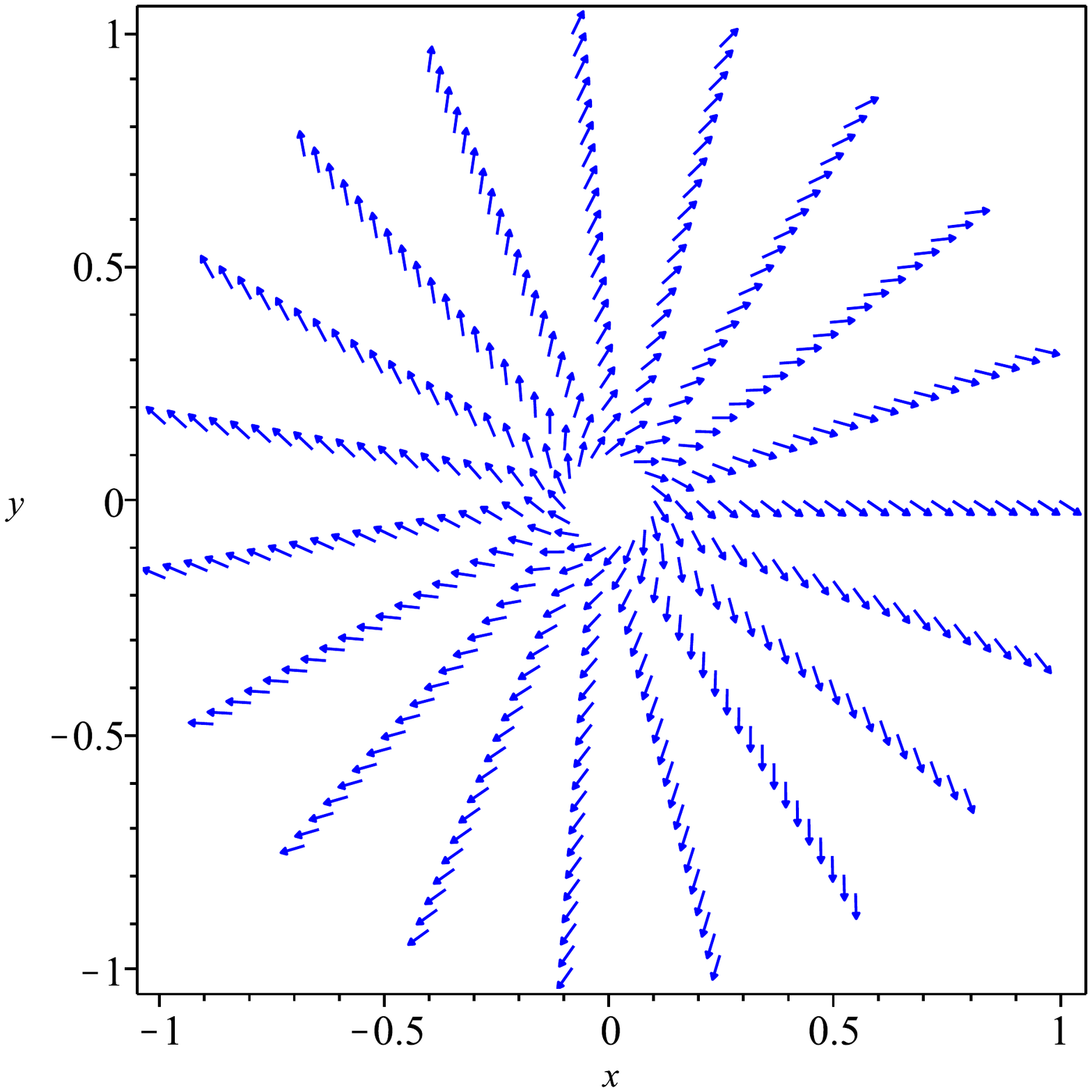}}}&
\rotatebox{0}{\scalebox{0.4} 
{\includegraphics{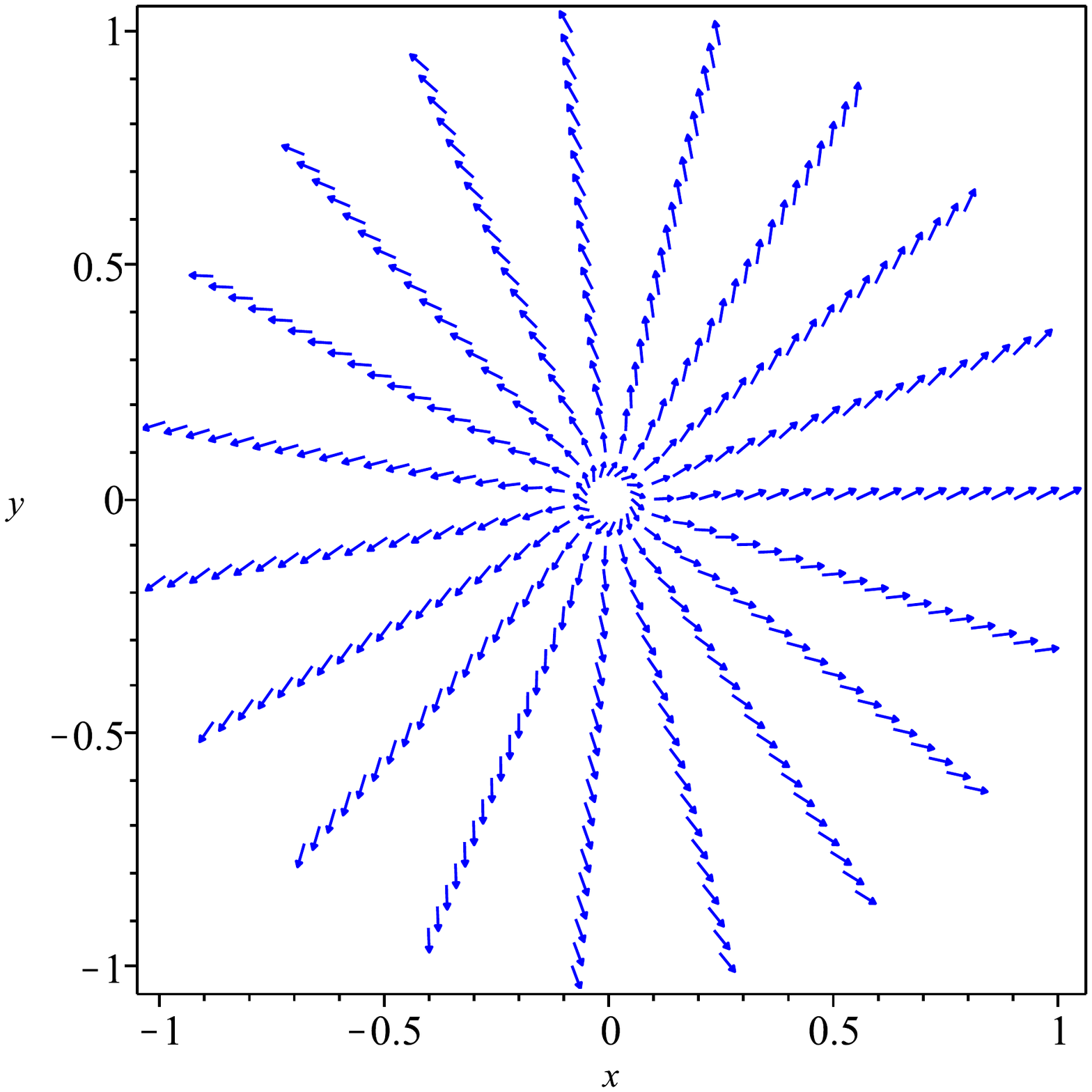}}}\\
{\rotatebox{0}{\scalebox{0.44} 
{\includegraphics{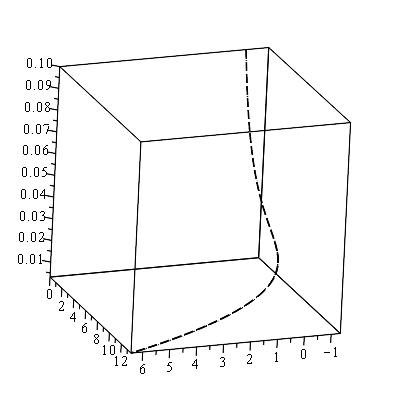}}}}&
\rotatebox{0}{\scalebox{0.44} 
{\includegraphics{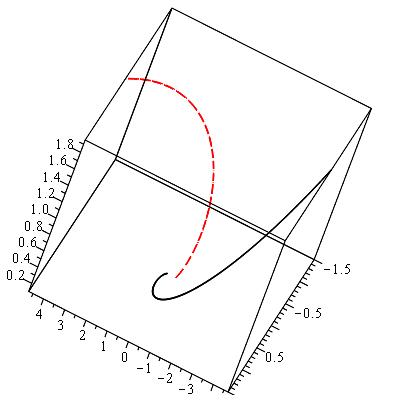}}}
\end{tabular}
\caption{At upper left we show the magnetic field vectors  in a cut at $z=0.05$. Using a  description parameter vector $[\tilde u_r,\bar u_z,s1,s2,s3,a,C_\phi,\Phi]$ , this is for the case $[-0.9,0.3,-1,-1,1,1,r,0.05,1,\pi/2]$ . At upper right we show the same cut for the same parameter vector 
except that $\Phi=1$. The radius runs from $0.1$ to $1$ at upper left and from $0.05$ to $1$ at upper right. At lower left we show a field line starting at   $[r,phi,z]=[0.5,\pi,0.1]$ for the same parameter vector as at upper left.   At lower right we show two field lines originating at $[r,phi,z]=[0.2,\pi,0.1]$ (solid) and 
 $[r,phi,z]=[0.2,0,0.1]$ (dashed). These have the same parameters as at upper right except that we have set $\bar u_z=-0.3$ (accretion) so as to remain inside our height limit and to accentuate the curvature. }
\label{fig:halofields}
\end{figure}

 In figure (\ref{fig:halofields}) the image at upper left shows the magnetic vectors in a cut at $z=0.05$. There is no sign reversal in each galactic quadrant. A typical field line is shown at lower left for this `quadrupolar' case. It descends rapidly to become parallel to the plane. At upper right  we show a  cut through a `dipolar' example at $z=0.05$ with the radius running from $0.05$ to $1$.  It is very similar to the similar cut for the exact dynamo solution in figure (\ref{fig:analytvphi}). A sign reversal in the first quadrant occurs at $r\sim 0.1$ ,where $Z=0.5$. The field lines at lower right are meant to show how this may occur.  Each field line is part of an {\it off axis} rising conical helix that is slowly turning. A field line initially directed towards us, turns away from the line of sight at large height. We have shown an accretion case $\bar u_z=-0.3$ in order to show the effect  within our height limit. The sub-scale Reynolds number $\Delta$ is $-0.822$ above the plane and $+0.822$ below the plane in our example.

 \begin{figure} 
   \centering
    \includegraphics[width=4in]{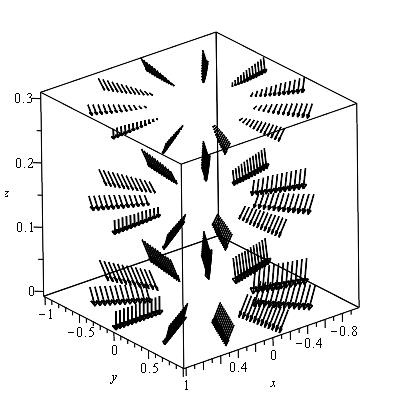} 
    \caption{ Using a parameter description vector $[\tilde u_r,\bar u_z, s1,s2,s3,a,C_\phi,\Phi]$, this case is given by $[ -0.9,0.3,-1,-1,1,1,-1,1]$ as at upper right in figure (\ref{fig:halofields}). In addition the radius runs over the range $[.05,1]$ and the height runs over the range $[0.01,0.3]$.  The figure shows the magnetic field vectors in three dimensions.   }
    \label{fig:dipfield3d}
 \end{figure}

 
 Figure (\ref{fig:dipfield3d}) illustrates the poloidal field associated with the toroidal magnetic field illustrated at upper right in figure (\ref{fig:halofields}). This field is pursued below for its rotation measure properties, but here it is clear that in projection it contains the `X' form of magnetic field and hence of polarization.
  
  As an amusing analogy, it is difficult not to remark on the similarity of these galactic magnetic fields to those of magnetic `skyrmions'  discovered in condensed matter physics (e.g. \cite{NT2013}). These are particularly prominent in thin films and the spiral or vortical spin skyrmions resemble those above.  Of course the spatial scale is in nano metres rather than in kiloparsecs!  Moreover the field is the actual `magnetic field' in the medium rather than the `magnetic induction' field used here. 
  
  \section{Rotation Measure}
  
  In the light of a recent discovery of oscillating rotation measure above and below the disc  due to rotation measure synthesis in the CHANGES galaxy NGC 4631 (\citet{CMP2017};\citet{CMP2016}; \citet{SPK2016}), we compute  the corresponding  rotation measure in the magnetic field of this example. We take the halo to have a cylindrical geometry with the galactic disc defining a section of the cylinder. The height is left undefined. 
  
  For these axially symmetric magnetic fields the rotation measure, integrated along the line of sight through the halo of a strictly edge-on galaxy, depends only on the azimuthal component of the field. For a galaxy inclination significantly different from $90^\circ$, the other field components will in general contribute. For simplicity here, where we are  only introducing  a possible oscillating effect in the rotation measure in each quadrant, we proceed with perpendicular  inclination. Moreover we treat the electron density $n_e$ as either constant or as an exponential in $Z=z/r$ . There is an arbitrary amplitude constant $C_\phi$ stemming from the magnetic field amplitude. 
  
  Under these conditions the RM, integrated along a line of sight  (los) through a section of the cylindrical halo, is given by (actually the negative of this quantity) 
  \be
  RM=C_\phi r_\perp \int_{-\sqrt{1-r_\perp^2/R^2}}^{\sqrt{1-r_\perp^2/R^2}}~n_e(r_\perp/\sqrt{(1-x^2)},z)\left(\frac{b_\phi(r_\perp/(\sqrt{(1-x^2}),z)}{1-x^2}\right)~dx,\label{eq:RMInt}
  \ee
  where $R$ is the radius of the disc and $x=\cos{\phi}$. The angle $\phi$ is between the los lying in a cylindrical section of the halo (the los has the impact parameter $r_\perp$ relative to the minor axis of the galaxy) and a line drawn to a point on the los  from the intersection  of  the galactic  minor axis with  the cylindrical cross section. 

Our calculation is  really the parallel magnetic field component integrated along the line of sight. As such it is the (negative)  of the Faraday rotation that would be produced by a Faraday Screen. One needs a model for the emitting electrons in the halo in order to calculate the RM quantitatively using radiative transfer, but we are  looking at qualitative possibilities here, so we hold it constant. The RM is weighted towards $x=0$ ($\phi=\pi/2$), so that its value reflects somewhat the  actual RM of a region near $r=r_\perp$.

\begin{figure} 
    \centering
    \includegraphics[width=4in]{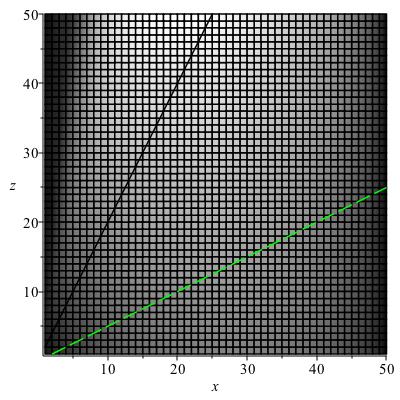} 
    \caption{The figure shows the  Faraday screen RM in the first quadrant of the quadrupolar field  $[-0.9,0.3,-1,-1,1,1,r,z,1,\pi/2]$ as in figure (\ref{fig:halofields}). The solid line shows on this grid the line $Z(r,z)=1$, because two units on the ordinate are equal to one unit on the abscissa . The dashed line is the locus of $Z(r,z)=0.5$.   }
    \label{fig:RMquadrupolar}
 \end{figure}

In figure (\ref{fig:RMquadrupolar}) we show a gray scale measure of this RM in the first quadrant for the quadrupolar field shown on the left hand side of figure (\ref{fig:halofields}).  The value varies from $+1.29$ at grid point $[25,30]$ to $[+0.89]$ at grid point $[25,5]$. There is smaller Faraday rotation near the axis, and the sign is the same everywhere in the quadrant. It would be the negative of this value on the other side of the galactic minor axis, although of course the signs may be interchanged. The region between the two lines on the figure is subject to  quantitative error but it seems qualitatively correct based on continuity.

 \newpage

\begin{figure}
\begin{tabular}{cc} 
\rotatebox{0}{\scalebox{0.4} 
{\includegraphics{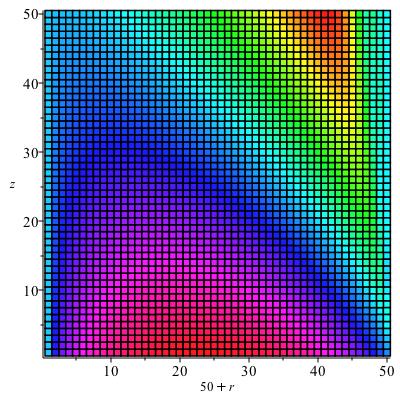}}}&
\rotatebox{0}{\scalebox{0.4} 
{\includegraphics{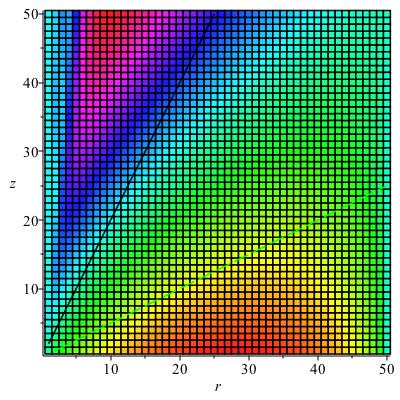}}}\\
{\rotatebox{0}{\scalebox{0.44} 
{\includegraphics{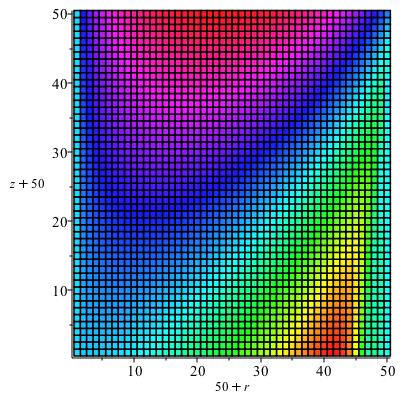}}}}&
\rotatebox{0}{\scalebox{0.44} 
{\includegraphics{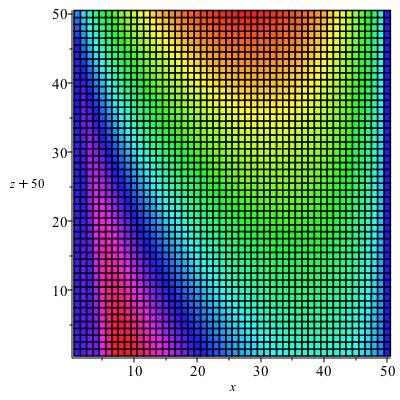}}}
\end{tabular}
\caption{ We have calculated the Faraday screen rotation measure for the magnetic fields having the parameter vector $[-0.9,0.3,-1,-1,1,1,r,z,-1,1]$ . The radius of the disc is equal to $1$. We have chosen vertical symmetry for the display, but diagonal symmetry is achievable by changing the sign of $C_\phi$ on crossing the plane. The upper line on the first quadrant is the locus $Z(r,z)=1$ and the lower line is the locus of $Z(r,z)=0.5$.    There is an arbitrary amplitude constant but the contrast between positive and negative RM is meaningful. In the first quadrant the orange peak ($[30,5]$ is negative at $~-0.67$ while the positive red peak $[10,40]$ has the value $~+0.65$.  The green shading $[30,25]$ is at $~-0.26$  while the blue shading at $[25,35]$ is at $~+0.3$.  }
\label{fig:RMquad1}
\end{figure}

\begin{figure} 
    \centering
    \includegraphics[width=4in]{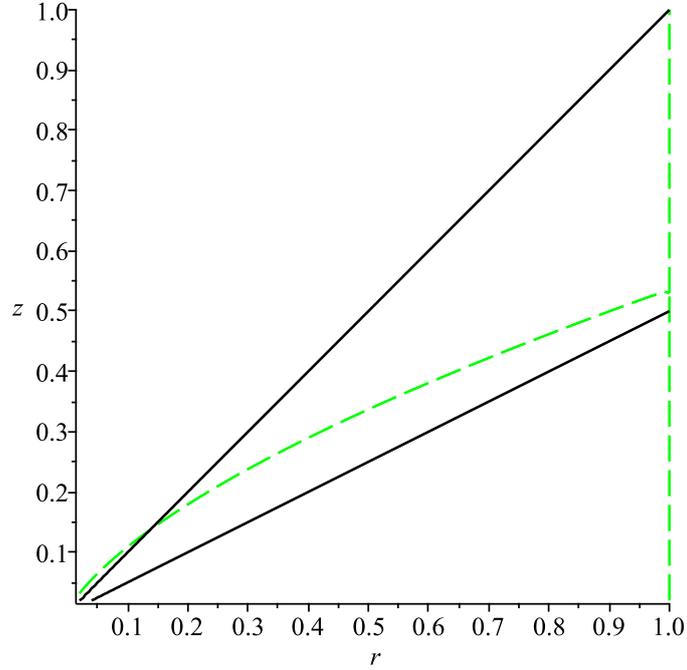} 
    \caption{The two solid lines correspond to $Z(r,z)=1$ and $Z(r,z)=0.5$. The curved dashed line shows the locus of the sign change in The RM as calculated for the distribution of figure (\ref{fig:RMquad1}).  The vertical dashed line is the edge of the disc.}
    \label{fig:Limits}
 \end{figure}

   In figure (\ref{fig:RMquad1})  we show the RM displayed over the halo for an edge-on galaxy having the parameter vector $[-0.9,0.3,-1,-1,1,1,r,z,1,1]$ , so that it is dipolar. The straight lines in the first quadrant are as in figure (\ref{fig:RMquadrupolar}).  The region between the straight lines is again subject to quantitative error. However qualitatively it shows the change in sign associated with the characteristic field lines at lower right in figure (\ref{fig:halofields}).  The curve along which the RM changes sign is shown relative to the two limiting straight lines is shown in figure (\ref{fig:Limits}). These lines are drawn in $\{r,z\}$ space so that $z=0.5$ corresponds to the top of figure (\ref{fig:RMquad1}). We see that the sign change region falls just within our height restriction. 
   
   The axes  of figure (\ref{fig:RMquad1}) should be interpreted in terms of the $ grid number/50 $ for the abscissa and $grid number/100$ for the ordinate.  The radius of the galaxy has been taken to be $50$ in grid Units ,that is  $1$ in  $\{r,z\}$ Units. The quadrants are readily formed by rotation from the first and second quadrants. However the fourth quadrant  may be calculated  directly , given the first quadrant, under the transformation $\{u,w,s1,s2,s3,a,r,z,C_\phi,\Phi\}\leftarrow\{u,-w,s1,-s2,-s3,a,r,z-0.501,C_\phi,\Phi\}$ . The number in the transformation implies calculation on a $50$ by  $50$ grid in steps of $0.02$ in radius and $0.01$ in height.  A similar transformation yields the third quadrant from the second quadrant by calculation, on using $r\leftarrow r-1.02$  and $z\leftarrow 0.501-z$ in addition to the other transformations.

   We have chosen to illustrate the case where $\bar b_\phi$ is symmetric on crossing the plane. If we had allowed $\bar b_\phi$ to change sign on crossing the plane than the symmetry would be `diagonal' rather than `vertical'.  This would be accomplished by changing the sign of $C_\phi$. A `butterfly' symmetry that would be horizontal is not physically reasonable in axial symmetry. The orange colour is negative according to our calculation, which would be measured as positive. Blue to red would then be measured as positive.

\begin{figure} 
    \centering
    \includegraphics[width=3in]{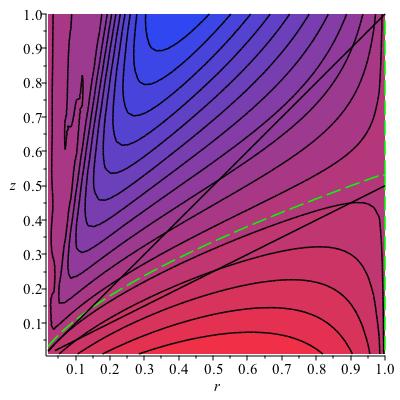} 
    \caption{Contours are overlaid on the RM distribution for the first quadrant in figure (\ref{fig:RMquad1}) of this section. The dashed line shows the zero contour. There are sixteen contours equally spaced between about $-0.67$ and $0.65$.  The right edge is the edge of the disc. The positive measured RM  (the negative of our values) peaks at about $0.6$ disc radii with the negative peak higher and closer to the axis. The upper solid line is the locus of $Z(r,z)=1$ while the lower solid line if that of $Z(r,z)=0.5$. }
   \label{fig:cquad}
 \end{figure}

 There is a conspicuous `X' type pattern to the RM distribution. This reflects the pattern expected in the linear polarization and in fact  known for some time (e.g. \cite{Kr2009}). The  coexistence of these two properties  may have been observed in (\cite{CMP2016}; \cite{SPK2016}).
  Figure (\ref{fig:cquad}) shows the RM distribution in the first quadrant with contours overlaid. The zero contour is shown as a dashed  line . The inclined nature of the  RM distribution is apparent.

\newpage

  As one might expect, the RM distribution is very sensitive to the assumed  electron density distribution. We have taken it constant until now to emphasize the sign changes in the RM {\it in each galactic quadrant } due to this dynamo magnetic field. For the illustrative parameters, the sign change occurs at considerable height in the halo. This makes the visibility of the sign change very sensitive to an exponential cut-off in the thermal electron density with height. The contrast with the region near the disc becomes extreme, even if  the absolute number may  still be detectable.  
  
   \begin{figure} 
     \centering
     \includegraphics[width=3in]{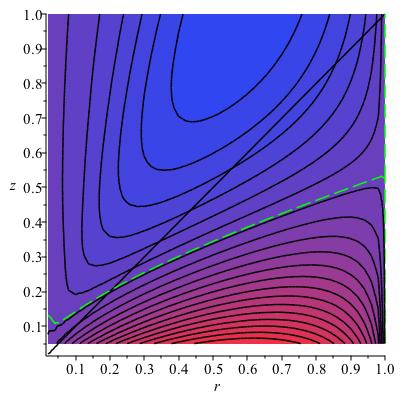} 
     \caption{ We show contours on the RM distribution of figure (\ref{fig:RMquad1}) in the first quadrant. The parameter vector is again  $[-0.9,0.3,-1,-1,1,1,r,z,-1,1]$, but the electron density has been taken as $\propto \exp{(-z/r)}$. The strength of the positive (mesured) RM region is  reduced. The upper solid line is the locus of $Z(r,z)=1$ while the lower solid line is that of $Z(r,z)=0.5$.}
     \label{fig:Qdip5}
  \end{figure}
  
  Applying density dependences in $r$ and $z$ separately will break the  scale invariance. One density variation that does not break the symmetry is 
  \be
  n_e=n_e(0)\exp{(- k Z)}\equiv n_e(0)\exp{(-kz/r)},\label{eq:SSdens}
  \ee
  where in the integration along the los we use $r=r_\perp/\sqrt{1-x^2}$.  This gives a variable scale height  equal to $r/k$. We show in figure (\ref{fig:Qdip5}) the case for $k=1$  for the same dynamo field as in figure(\ref{fig:RMquad1}). The dashed line is the zero line  and the solid line is $Z(r,z)=1$. Further discussion along these lines must await a more extensive parameter search motivated by data.

  \newpage 
 \section{Discussion and Conclusions}  
   
   One can hardly imagine a simpler origin of a galactic magnetic field than a scale-free, axially symmetric, steady, mean field dynamo as presented here\footnote{Fossil magnetic fields have been suggested and were considered already in the sixties and seventies as involved in galaxy formation and activity--see e.g. \cite{HR1977}}. The number of parameters are reduced to a bare minimum. 
   
   Clearly it has not been my intention to advance dynamo theory itself, but rather to apply classical theory to a developing set of observations ( CHANG-ES, \citet{WI2015}). Mean field dynamo theory, particularly in numerical simulations \cite{B2014},\cite{KF2015}, has indeed  advanced well beyond the classical theory. However the `no z' approximation (e.g. \citet{B2014},\cite{M2015}) has led to the dynamo structure above the spiral disc being somewhat neglected recently.
   
   **
   
  However a referee has  quite properly called my attention to earlier work that is relevant to the results of this paper. This concerns the extended discussion in the literature  of the possibility of different `parity' relative to the galactic equator between the equatorial regions and the halo. This suggestion appears to be found originally in \cite{SS1990}, wherein  the authors observe that this may be true due to different geometries. This suggestion was confirmed in \cite{BDMSST92}. {\it Such a parity contrast  implies the `sign change' in the azimuthal magnetic field that we have found here}. The calculation was time dependent and assumed rather particular (however physically motivated) variations in diffusivity and helicity. Their results also exhibited sensitivity to poorly known initial conditions, and a steady state was not easily attained.   
  
  In the present study, the assumption of scale invariance  and the steady state has allowed us to dispense with particular assumptions about the diffusivity and helicity, so long as they are proportional on cones. Nevertheless, this steady state effect could be described as a `parity oscillation' in each quadrant as in the earlier papers.  
  
 Motivated by observations of the Milky Way  suggesting this effect  \cite{sun2008}, this possibility was studied again in a generic fashion in 
 \cite{MS2008}. These authors studied once again time dependent solutions of the classic dynamo equations  with particular choices of diffusivity and helicity. They  found the desired effect (their model 134b) {\it only} in an oscillatory  model for a careful choice of parameters. For this reason their principal conclusion was negative, although they do remark (without elaboration) that this effect may be found in steady solutions. 
 
   Our intention in this paper was to produce a semi-analytic model of likely steady magnetic dynamo structure above the disc that may be easily reproduced and compared to observations.  Our results depend on the self-similar assumption and the steady state, whereas earlier work introduced time dependence and particular distributions of diffusivity and helicity. To obtain a steady state after temporal evolution some form of `quenching' is required. Such an effect is potentially present in our approach since a certain velocity structure is generally required for each of our solutions. Nevertheless  despite the marked differences in approach, the apparent convergence of this work with earlier work comprises support for both techniques.
   
     This paper presents the axially symmetric model (the same symmetry as in the earlier studies referred to above) and a companion paper  \cite{Hen2017} discusses magnetic spiral arms.  The detailed  comparison with observation must await a subsequent paper.
     
 **  
 
  The most relevant  insight from this work is that the  dynamo with globally constant sub-scale Reynolds number  (this is essentially the `Dynamo number' as defined in \citet{B2014}) can produce helical magnetic fields on cones that  are `off axis'.  That is, the axis of the helical cone does not coincide with the minor  axis of the galaxy.  This behaviour is not  sensitive to variation in the helicity or the diffusivity on cones, so long as these vary proportionally on cones. The solutions thus contain in principle a general form of gradients in the helicity and diffusivity (see e.g. \citet{M2015}) . The twist of these fields varies with radius at a given height or vice versa, due to the scale invariant symmetry.  This leads to sign
 reversals in the azimuthal magnetic field  and consequently of the RM in each quadrant of the edge-on galaxy in the sky plane. This is not possible in axial symmetry if the axes of the field lines coincide with the galactic axis.   
 
 This single quadrant sign reversal of the azimuthal magnetic field (i.e. `parity oscillation')  is reflected in the  Faraday screen Rotation Measure.  Indications of such behaviour have recently been discovered (\cite{CMP2016}; \cite{SPK2016};\cite{CMP2017}) in the CHANG-ES survey galaxies, but  detailed comparisons will be necessary . These must include the addition of asymmetric modes. Ultimately radiative transfer through a realistic model halo must be carried out to interpret the results of rotation measure synthesis (e.g \citet{HF2014}).

 The model discussed mostly here is always solenoidal (as it follows from the equations for the vector potential) but it is limited in accuracy in the vertical direction, especially at small radius. This has been discussed at length in section 4.  However even with $Z=0.2$, we are justified in reaching one kilo parsec into the halo at a radius of five kilo parsecs. This is well into the `halo' as defined by the edge-on galaxy observers. Although subject to quantitative error, it does seem that the sign change effect is real and comprehensible. 
 
 We have studied the  equations (\ref{eq:beqs}) for simple  (i.e. `toy' models) solutions that apply at all heights in the galactic halo in order to justify the approximate treatment. {\it Both solenoidal and non-solenoidal solutions of these classic mean field dynamo equations exist}.  These solutions hold at all heights in the halo and their behaviour is similar to that studied approximately. These solutions have allowed us to explore other scale invariant `classes' such as $a=2$ and $a=3$ and to discuss their implications. They all contain the sign-changing effect in one quadrant of an edge-on galaxy.
 
 One should recall equations (\ref{eq:Scinv}) for the radial dependences of fields, helicity and diffusivity.  Appendix  A shows how these may be generalized by the addition of parallel and perpendicular scale lengths. When $a=3$ the solenoidal solution of equations (\ref{eq:beqs}) exhibits very similar sign changing  behaviour to that shown in our approximate cases. However it is not so readily calculated as either the non solenoidal example or our approximate solution (being dependent on Legendre functions), so that we display the simpler cases.
 
 **
 
 The pure $\alpha-\omega$ dynamo example ( $\Delta=0$) is  solenoidal. It does not have the correct X type projection, but it does include the principal result of this paper for any value of $Z$. That is, that axially symmetric steady dynamos may have different signs of rotation measure in each galactic quadrant. It is not however a solution with which we would fit to observational data. Either our  general explicit forms or numerical solutions to the self-similar equations will permit this in a subsequent study.
 

 
 The question as to the physical significance of the non-solenoidal equations (\ref{eq:beqs}) remains. We have allowed ourselves a speculation that is now presented in an appendix.  Briefly we  suggest that a non solenoidal mean field implies a different averaging volume over different regions of the sub-scale magnetic field. The steady state  mean magnetic field equations do not incorporate the solenoidal requirement directly (e.g. \citet{M1978}). Even the time dependent equations must start from this initial condition and then maintain it rigorously in order to arrive at a solenoidal state. It seems to us that different regions of a turbulent dynamo may have to be averaged over different volumes in order to reflect a constant helicity and diffusivity on cones. Different sized regions may even evolve to a steady state at different times. In any case, this speculation is not essential to the main result of this paper. 
 A demonstration that equations (\ref{eq:beqs}) may be derived independently of the solenoidal condition is also given in that Appendix, which is quite independent of the speculation.
 
 **
  
 At present only very simple electron density distributions (constant or exponential in $z/r$) have been considered. Moreover the halo geometry is cylindrical and the inclination of the galaxy is taken as $90^\circ$. Modifying these assumptions adds parameters that should be considered in making comparisons with the observations. For simplicity while presenting our main effect qualitatively we have avoided these in the present paper. 
 
 It is already clear that the field can produce Rotation Measure distributions that are either symmetric about the galactic plane or are diagonally symmetric   through the galactic centre while always maintaining an X type pattern in each quadrant.  A distribution with `butterfly symmetry', that is horizontal symmetry, can not be plausibly produced as that would require changing the field sign across the galactic minor axis.   In order that RM distributions may  reveal the galactic plane magnetic field, the handedness of the magnetic field  should not change on crossing the plane. That is, the azimuthal field should be constant through the plane.  
 

 Our parametric method avoids the difficult problem of solving simultaneously for the dynamics and the dynamo  magnetic field either by using the scaled velocity as a parameter or by assuming parallel velocity and magnetic fields \cite{HI2016}. However the combination of scale invariance and axial symmetry leads to an over determined problem for the vector potential since the solution is also constrained to be solenoidal. This can be resolved through restrictions on the parameters, but the solution is not unique.  It is possible to let vary  $\alpha_d$, $\bar\eta$ and $\bar v$ , all as the same arbitrary function of $Z$, and still obtain our results. These freedoms do not change the solution for the reduced (constant) velocity $u$ and the scaled magnetic field, but it does increase the generality of the physical velocity field and allows the dynamo parameters to vary across cones (cf \citet{M2015}). Should we allow our sub-scale Reynolds number parameter (i.e essentially the Dynamo number $\Delta$) to be a function of $Z$ to be found as part of the solution, the problem ceases to be overdetermined but becomes strongly non-linear. Nevertheless this provides a method to proceed numerically.
 
 The scale invariant equations (\ref{eq:beqs}) are not over determined since they are not constrained to be solenoidal. They are exact and could be studied at length numerically. We have been content to find some exact analytic solutions both solenoidal and non solenoidal.
 A more numerical study will be important when observational data are to be fitted, although trends should be clear from our results.  The necessary data is still developing \cite{CMP2017}. A sign changing effect is of course to be expected if the magnetic spirals \cite{Beck2015} persist into the halo. A companion paper  \cite{Hen2017} discusses that possibility using the same methods. Nevertheless it is important to have such an effect present in axial symmetry since this may be the ultimate form of the magnetic field if it is subject to differential rotation \cite{B2014}. This is the same concern that appears when discussing the material arms. 
 
 Despite the limitations of our exploratory study, the magnetic fields shown in our figures contain many observed properties. One can reproduce the `X-type' polarization structure, horizontal or vertical RM sign changes, and hints of RM sign variation in each quadrant in the halo of an edge-on galaxy.  The magnetic field behaviour is rather similar to that suggested for the Milky Way \cite{Gfarr2015} and inferred empirically \cite{FT2014}.

   \section {Acknowledgements}
   
   This study was achieved while the author was enjoying the hospitality of the Astronomy group at the R\"uhr Universit\"at Bochum. The stay was made possible  by a generous award from the Alexander von Humboldt foundation together with the welcome provided by professor Ralf-Juergen Dettmar and members of his group. I wish to thank Reiner Beck for discussion and helpful criticism of an earlier draft of this paper.  Marita Krause, Carolina Mora, Philip Schmidt and Arpad Miskolczi are to be thanked for discussions. Dr. Judith Irwin always advises wisely. A referee has helped to sharpen the arguments. 
   
   \section{Appendix A: Generalized Power law behaviour}
   
  The multiplicative  radial dependences  of our solutions  are always a simple power law, depending on the class parameter $a\equiv \alpha/\delta$. We have used  a mean isotropic spatial scaling $\delta$ in the text. It is possible to generalize the radial dependence slightly in terms  of distinct parallel  $\delta_\parallel$ and perpendicular $\delta_\perp$ reciprocal scales, which device yields slightly more flexible radial power laws ($1/\delta_\perp$ would be related to a scale height). In that case the form of the scale invariance becomes ($R$ and $Z$ are as in the text)
\bea
\mt {A}&=&{\bf \bar A}(Z) e^{(2\delta-\alpha)R},\nonumber \\
\mt{v}&=&{\bf \bar v}(Z) e^{(\delta_\perp-\alpha)R},\label{eq:altScinv}\\
\alpha_d&=& \bar\alpha_d(Z) e^{(\delta_\perp-\alpha)},\nonumber\\
\eta &=& \bar\eta(Z) e^{(\delta+\delta_\perp -\alpha)},\nonumber
\eea
 and now 
 \be
~~~~~~~~~~~~~~~~~~~~~~~~~~~~ ~~~~~~~~~~~~~~\delta=\frac{\delta_\parallel+\delta_\perp}{2}\label{eq:avgdelta}.
 \ee

This treatment of scale invariance follows from the analysis advocated in (\cite{CH1991} )and in detail in \cite{Hen2015}) . It is similar in spirit to the two dimensional scaling used in the Blasius solution for viscous flow over a  flat plate. It allows us to use different scalings in the directions perpendicular and parallel to the plane. 

 We have scaled the sub-scale helicity and the  mean velocity (in whatever pattern frame)  in terms of the perpendicular scale height, because this corresponds to rotating gas rising  from the galactic plane.  We scaled the vector potential in terms of the isotropic reciprocal scale, assuming it to have similar characteristic scales in both directions.
 The innovative  scaling  is that of the  diffusion coefficient. It must have a net  spatial scaling (the dependence through $\delta$ as in  the text equation (\ref{eq:altScinv})) of two Dimensionally, and a Dimensional temporal scaling of one (the dependence through $\alpha$). We have achieved this by using one perpendicular reciprocal scale plus the average reciprocal scale. This choice is necessary to create scale invariance given our other choices. However we can expect the diffusion to involve both perpendicular and parallel scales in fact because it is likely turbulence dependent.  We note that this requires the two spatial reciprocal scales to be divided according to $3\delta_\perp/2+\delta_\parallel/2$, so that the perpendicular scale is weighted more heavily.   However, it is possible to exchange $\delta_\perp$ for $\delta_\parallel$ throughout, if this were believed to be more physically reasonable.
 
We are free to define the `similarity class'  once again as $a\equiv \alpha/\delta$.  Hence the radial dependences of the velocity , the sub-scale helicity and the diffusion coefficient become (recall  the text equation (\ref{eq:rz}) for $\delta r$)
\bea
\mt{v},\alpha_d &\propto& (\delta r)^{(\frac{2\delta_\perp}{\delta_\parallel+\delta_\perp}-a)}  ,\nonumber\\
\eta&\propto&(\delta r)^{(\frac{3\delta_\perp}{2\delta}+\frac{\delta_\parallel}{2\delta}-a)}.\label{eq:radial}
\eea

The standard procedure sets $\delta_\perp=\delta_\parallel=\delta$, so that these dependences become $(1-a)$ and $(2-a)$ respectively, as in equations (\ref{eq:Scinv}) of the text.  However, these dependences depend now on the relative size of the two directional reciprocal scales.  For example if $\delta_\perp\gg\delta_\parallel$ then the powers tend to $2-a$ and $3-a$ respectively.  If $\delta_\parallel\gg\delta_\perp$ then we obtain $-a$ and $1-a$ for the respective powers. The class $a$ may also change depending on how a governing integral acts. {\it Because of the integrated form of equation (\ref{eq:dynamoI}), the equations for the vector potential are not different from the standard case obtained when the scales are equal, as in the text}. We can therefore invoke this optional scaling as desired. 

**

 \section{Appendix B: Steady State dynamo; Solenoidal Mean Magnetic field?}
 
 The solenoidal/a-solenoidal ambiguity appears to be because the  `ab initio' {\it steady state} dynamo theory is ambiguous in this regard.  Thus consider that  we expect  from Ohm's law  that the total {\it steady}  electric field ${\bf E'}$ in the medium moving with velocity ${\bf v}$ is 
 \be
 c{\bf E'} =\frac{{\bf j}}{\sigma},
 \ee
 where ${\bf j}$ is the current density and $\sigma$ is the  effective conductivity.  Consequently from Faraday's law and Amp\`ere's law (and $v\ll c$) in the absence of a conservative electric field, we have in the systemic frame the total steady electric field as 
 \be
 c{\bf E}=-{\bf v}\wedge\vec {\bf B}+\eta\nabla\vec {\bf B}={\bf 0},\label{eq:Ezero}
 \ee
 where ${\bf B}$ and ${\bf v}$ are total quantities, $\eta$ is the sub scale resistivity, and the derivatives are over a sub-scale region.
 Normally one averages this equation over a  constant mesoscale volume after writing the velocity and magnetic fields in terms of mean and fluctuating parts. Making the usual assumptions about the sub-scale behaviour (\cite{M1978}), gives {\it directly} for the mean magnetic field ${\bf b}$ 
 \be
 \alpha_d {\bf b}-\eta\nabla\wedge {\bf b}+{\bf v}\wedge {\bf b}={\bf 0}.\label{eq:steadyBdynamo}
 \ee
 This becomes our equations \ref{eq:beqs} after introducing self-similarity and writing out the components. We do not require from the formalism that the mean field be solenoidal to this point. 
 
 As a speculation concerning the relevance of non solenoidal mean magnetic dynamo fields, we offer the following.
 The mesoscale averaging volume is usually assumed to be constant (ensuring that it commutes with differential operators) and hence the mean magnetic field should be solenoidal if the sub-scale  magnetic field is solenoidal. However, provided that  the mesoscale average over the  sub-scale curl of the mean magnetic field may be taken equal to the macroscopic curl of the mean magnetic field, one obtains the same equation for the mean field ${\bf B}$ even if the mesoscale averaging volume is not constant.  The solenoidal constraint is not then necessarily present.  
 
 This  debatable condition should be examined.  It takes the  explicit form  
 \be
 \bar\nabla\wedge {\bf b}=\big<\nabla\wedge {\bf b}\big >,
 \ee
where  $\bar\nabla$ refers to macroscopic derivatives, $\nabla$ refers to sub scale derivatives and ${\bf b}$ is the mesoscale averaged magnetic field. The indicated average is the mesoscale average. If expressed in Cartesian component form we have 
\be
D_jb_k=\big<\partial_jb_k\big>,
\ee
where $D_j$ is the macroscopic derivative and $\partial_j$ is the sub-scale derivative. Because there is no other derivative in equation (\ref{eq:Ezero}), this may actually be used as the meaning of the macroscopic derivative, when acting on any  function  that is averaged over randomly fluctuating values .  
 
**

\label{lastpage}

\newpage


\begin{thebibliography}{99}
\bibitem[\protect\citeauthoryear{Beck}{2015}]{Beck2015} Beck, Rainer, 2015, Astronomy \& Astrophysics Review, {24},{4}
\bibitem[\protect\citeauthoryear{Blackman}{2015}]{Black2015} Blackman, E. G., 2015, Space Science Reviews, {188}, {59}
\bibitem[\protect\citeauthoryear{Brandenburg et al.}{1992}]{BDMSST92} Brandenburg, A., Donnor, K. J., Moss, D., Shukurov, A., Sokoloff, D., Tuominen, I.,
1992,\aap,{259},{453}
\bibitem[\protect\citeauthoryear{Brandenburg}{2014}]{B2014} Brandenburg, A., 2014, {\it Simulations of Galactic Dynamos} in {\it Magnetic Fields in Diffuse Media}, \# 407, {\it Astrophysics and Space Science Library}, 529
\bibitem[\protect\citeauthoryear{Brentjens\& de Bruyn}{2005}]{BdeB2005}  Brentjens, M.A. \& de Bruyn, A.G., 2005, \aap, {441},{1217}
\bibitem[\protect\citeauthoryear{Carter\&Henriksen}{1991}]{CH1991} Carter, B. \& Henriksen R. N., 1991, J. Math. Phys., {32(10)},{2580}
\bibitem[\protect\citeauthoryear{Damas-Segovia et al.}{2016}]{DamSegov2016} Damas-Segovia, A., et al., CHANG-ES VII, 2016, \apj, {824}, {30}
\bibitem[\protect\citeauthoryear{Farrar}{2015}]{Gfarr2015} Farrar, G. 2015, {\it Astronomy in Focus}, vol. 1, XXIXth IAU General Assembly, Ed. Piero Benvenuti
\bibitem[\protect\citeauthoryear{Ferri\`ere \& Terral}{2014}]{FT2014} Ferri\`ere, K., \& Terral, P. 2014, \aap, {561}, {100}
\bibitem[\protect\citeauthoryear{Heald et al.}{2007}]{H2007} Heald, G., Rand, R., Benjamin, R., \& Bershady, M. 2007, \apj, {663}, {933}
\bibitem[\protect\citeauthoryear{Heald}{2009}]{Heald2009} Heald, G., 2009, {\it Cosmic Magnetic fields}, IAU Symposium, {259}, {591}, Strassmeier, K.G., Kosovichev,A.G. \& Beckman, J.E., Eds 
\bibitem[\protect\citeauthoryear{Henriksen\&Reinhardt}{1977}]{HR1977} Henriksen, R.N. \& Reinhardt, M.,1977, Ap\&SS, {49},{3}
\bibitem[\protect\citeauthoryear{Henriksen}{2015}]{Hen2015} Henriksen, R. N., 2015, {\it Scale Invariance: Self-Similarity of the Physical World}, Wiley-VCH, 69469 Weinheim,Germany
\bibitem[\protect\citeauthoryear{Henriksen\&Irwin}{2016}]{HI2016} Henriksen, R.N. \& Irwin, J.A., 2015,\mnras,{458},{4210}
\bibitem[\protect\citeauthoryear{Henriksen}{2017}]{Hen2017} Henriksen, R.N., \mnras, submitted
\bibitem[\protect\citeauthoryear{Horellou\&Fletcher}{2014}]{HF2014} Horellou, Cathy \& Fletcher, A., 2014,\mnras,{441},{2049}
\bibitem[\protect\citeauthoryear{Klein\& Fletcher}{2015}]{KF2015} Klein, U. \& Fletcher, A., 2015,{\it Galactic and Intergalactic Magnetic Fields}, Springer,
Switzerland
(
\bibitem[\protect\citeauthoryear{Krause}{2009}]{Kr2009} Krause, M. 2009, Rev. Mex. AA, {36}, {25}
\bibitem[\protect\citeauthoryear{Krause}{2015}]{Kr2015} Krause, M. 2015, {\it Highlights of Astronomy}, {16}, {399}
\bibitem[\protect\citeauthoryear{Nagaosa\&Tokura}{2013}]{NT2013} Nagaosa, N. \& Tokura, Y., Nature Nanotechnology, {8},{899}
\bibitem[\protect\citeauthoryear{Moffat}{1978}]{M1978} Moffat, H. K. 1978, {\it Magnetic field generation in electrically conducting fluids}, Cambridge University Press, Cambridge, U.K.
\bibitem[\protect\citeauthoryear{Mora Partiarryo}{2016}]{CMP2016} Mora Partiarroyo, S. C., Ph.D. Thesis, Bonn Universit\"at, Bonn, Germany, http://hss.ulb.uni-bonn.de/2016/4537.htm
\bibitem[\protect\citeauthoryear{Mora Partiarryo}{2017}]{CMP2017} Mora Partiarroyo, S. C., In preparation
\bibitem[\protect\citeauthoryear{Moss\&Sokoloff}{2008}]{MS2008} Moss, D. \& Sokoloff,D. 2008, \aap, {487},{197}
\bibitem[\protect\citeauthoryear{Moss et al.}{2015}]{M2015} Moss, D. Stepanov, R., Krause, M.,Beck, R., Sokolloff,D., 2015, \aap,{578}, {94}
\bibitem[\protect\citeauthoryear{Rand}{2000}]{R2000} Rand, R. 2000, \apj, {537}, {L13}
\bibitem[\protect\citeauthoryear{Schmidt, Partiarroyo\& Krause}{2016}] {SPK2016}Schmidt, P.,Partiarroyo, S.C.M. \& Krause, M. 2016, Final Annual Meeting of the DFG research unit 1254, "Magnetization of Interstellar and Intergalactic Media",Berlin, October
\bibitem[\protect\citeauthoryear{Sokoloff\&Shukurov}{1990}]{SS1990} Sokoloff, D. \& Shukurov, A., 1990, Nature, {347},{51}
\bibitem[\protect\citeauthoryear{Sun}{2008}] {sun2008} Sun, X. H. 2008, \aap, {477}, {573}
\bibitem[\protect\citeauthoryear{Wiegert et al.}{2015}]{WI2015}Wiegert, T., Irwin, J. A., Miskolczi, A., Schmidt, P., Mora, S. C., Damas-Segovia, A., Stein, Y., English, J., Rand, R. J., Santistevan, I., plus 14 co-authors 2015, \aj, {150}, {81}
\end{thebibliography}
\end{document}